# Tilt or twist – competing synclinic and anticlinic interactions in SmC phases of bent-core mesogens

Jiří Svoboda[1], Václav Kozmík[1], Kvetoslava Bajzíková[1], Michal Kohout[1],

Vladimíra Novotná[2], Natalia Podoliak[2], Damian Pociecha[3], and Ewa Gorecka[3]

[1] *Department of Organic Chemistry, University of Chemistry and Technology, CZ-166 28 Prague 6, Czech Republic*

[2] *Institute of Physics of the Czech Academy of Sciences, Na Slovance 2, CZ-182 21 Prague 8, Czech Republic*

[3] *Chemical Faculty, Warsaw University, Al. Zwirki i Wigury 101, 02-089 Warsaw, Poland*

Abstract:

Recent liquid-crystalline (LC) research is focused on structurally new molecular systems distinct from simple nematic or smectic phases. Sophisticated molecular shape may reveal structural complexity, combining helicity and polarity. Achiral symmetry-breaking in bent-core molecules leads to propensity for synclinic and anticlinic molecular structures within consecutive smectic layers. Moreover, despite their achiral character, dimers readily adopt helical phases. In our study, we investigated a hybrid molecular structure incorporating both characteristics, namely a rigid-bent core and an attached bulky polar group via a flexible spacer. To perform phase identification, we enriched the standard experimental methods with the sophisticated resonant soft x-ray scattering. Notably, we have observed a distinct preference for specific phase types depending on the length of the homologue. Longer homologues exhibit a predisposition towards the formation of tilted smectic phases, characterized by complex sequences of synclinic and anticlinic interfaces. Conversely, shorter homologues manifest a propensity for helical smectic structures. For intermediate homologues, the frustration is alleviated through the formation of several modulated smectic phases. Based on the presented research, we describe the preconditions for high level structures in relation with conflicting constraints.

Key-words: bent-core liquid crystals; terminal nitrophenyl unit; mesomorphic properties;

## 1. Introduction

For years, two categories of thermotropic liquid crystals have captured attention: those composed of rod-like molecules and forming nematic and smectic phases, and disc-like molecules preferring columnar phases [1]. In last years, the focus switched to less conventional mesogenic structures, the group that became particularly intriguing being the molecules with a bent shape [2-9]. These molecules are either dimers, composed of two mesogenic parts connected with a flexible spacer with an odd number of atoms [10-14], or they have an extended rigid core with a central usually meta-substituted phenyl or naphthalene ring [15-23]. A rigid bent-core shape restricts the molecular rotation, which in turn might induce the correlations of dipoles. Bend dimers also facilitate low bend elasticity, which leads to heliconical molecular arrangement [24]. As a result, while bent rigid core mesogens have a strong tendency to form polar phases [14-16], flexible dimers readily form helical phases [25]. In this study, our focus is on the molecules that combine both characteristics. Studied compounds feature a large rigid core typical for bent-core molecules, prolonged in one arm via a flexible alkoxy linker, which is terminated by bulky nitrophenyl moiety.

In our previous work [26], we studied the effect of various terminal groups (phenylalkyl, phenyloxyalkyl, thienylalkyl, and substituted phenylalkoxy) on the mesomorphic properties of a series of naphthalene-based bent-core molecules. For the materials presented here, we have modified the polarity of the terminal group in comparison with the previously introduced molecular structures [27-29] to enhance dipolar interactions. Our aim is to shed light on the modification of self-assembly processes in bent core molecules with bulky polar end group attached with a flexible spacer.

## 2.    Results and Discussion

A series of mesogens with molecules having rigid naphthalene-based bent core, terminated on one side with dodecyloxy chain, has been synthesized (Figure 1). A nitro-substituted phenyl ring is attached to the core via a flexible alkoxy linker on the other side. The homologues are denoted as **I-n**, where *n* represents the number of carbon atoms in the spacer, ranging from 4 to 12. The synthesis procedures are described in Supporting Information (SI). The phase transition temperatures and corresponding enthalpy changes, obtained from DSC measurements, are summarized in Table S1 and Figure S1 in SI for each homologue, and the final phase diagram is presented in Figure 2. The phase transitions were clearly distinguishable during DSC measurements. The only exception is the $SmC_1$- $SmC_2$ phase transitions, which were detected based on the textural changes and birefringence measurements. The clearing, as well as the N-SmA phase transition temperatures exhibit a general decrease on elongation of the molecules, an odd-even effect is superimposed on this trend, which is especially pronounced for short homologues. Molecules having a spacer built of an even number of atoms display higher transition temperatures, which is attributed to a more efficient packing in the case when the terminal nitrophenyl group is aligned parallel to the mesogenic arm of the molecule.

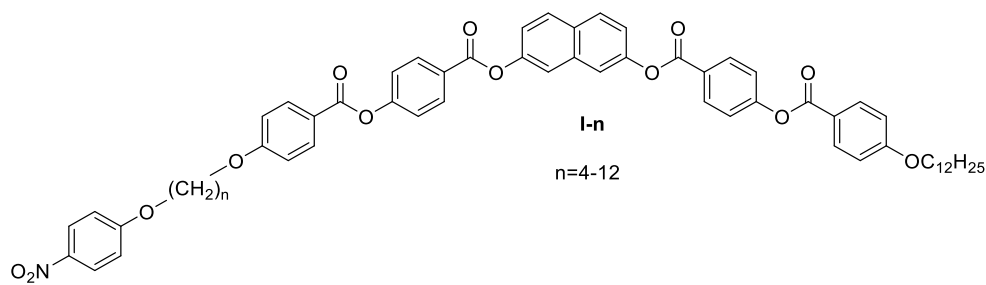

**Figure 1.**    General chemical formula of the studied compounds.

Except for the materials with the shortest alkyl chain (**I-4** and **I-5**), all compounds exhibit a nematic phase below the isotropic liquid (Iso), which is followed by a smectic A phase on cooling. Contrary to the behaviour of typical rod-like mesogens, for which elongated terminal chains destabilise the nematic phase in favour of the smectic phases, in studied homologue series the nematic phase temperature range expands with the elongation of the alkoxy spacer group. Most probably, when the lengths of alkyl chains attached to both arms of the mesogenic cores becomes comparable, the interactions between the mesogenic cores and the nitrophenyl terminal groups increase, which reduces the tendency to form layers.

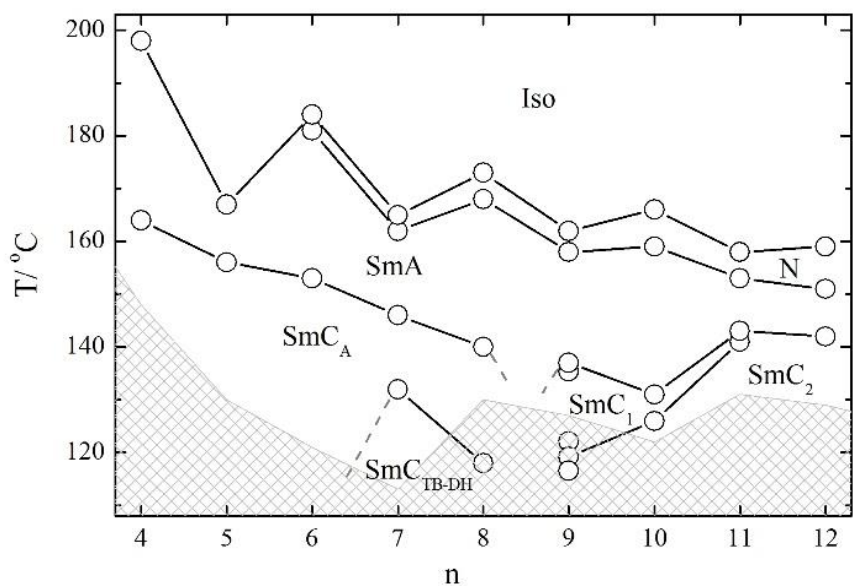

**Figure 2**. Phase diagram for the studied homologues upon elongation of the spacer between the rigid core and the terminal nitrophenyl group. Gray tilled area shows the stability range of the crystal phase marked by melting points.

A phase diagram of series **I-n** (Figure 2) is essentially split into two distinct regions, with homologue **n**=9 possessing a crossover behaviour. For homologues with **n**<9, tilted phases, either anticlinic or heliconical (of the $SmC_{TB}$ type), are formed below the SmA phase. On the other hand, for homologues with **n**>9, non-helical tilted smectic phases are observed, displaying a complex tilt structure. This discrimination is also evident when analysing the temperature dependence of the layer spacing, $d$, in smectic phases (Figure 3). Short homologues exhibit positive thermal expansion of the layer thickness in the whole temperature range and a very strong odd-even effect – layer spacing is systematically larger for even homologues than for the neighbouring odd ones. In contrast, longer homologues in lower temperature phases display negative thermal expansion of the layer thickness, and the parity of the spacer has much weaker effect on the $d$ value.

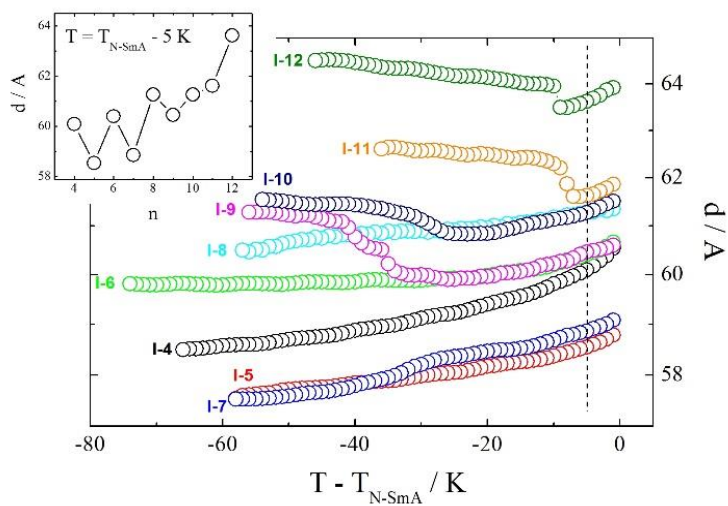

**Figure 3**. Smectic layer thickness, *d*, measured as a function of temperature for the studied homologues **I-n**. In the inset – the layer spacing in the SmA phase, determined 5K below Iso-SmA or N-SmA phase transition (dashed line on main panel) is presented.

### 2.1. Mesomorphic properties of the short homologues

In this section, we will describe the behaviour of the short homologues, with n<9. Below the Iso and N phases, the SmA phase appeared. The SmA phase is optically uniaxial and in the cells with homeotropic anchoring (HT) a uniform black texture without any recognisable features is observed. Upon cooling to the phase below the SmA phase, a schlieren-like texture emerges (Figure S2), which could suggest either a $SmA_b$ (biaxial non-tilted smectic phase, in which molecules show restricted rotation along the long axis) or a tilted smectic phase. Doping of the studied material with a small amount of chiral additive makes the phase optically uniaxial, which allows for the identification of a tilted, SmC-type, phase. For homologues **I-7** and **I-8**, another SmC phase is observed on further cooling, having uniaxial optical properties (appearing dark in the case of homeotropic anchoring when placed between crossed polarisers, see Figure S3). The X-ray diffraction studies revealed in all smectic phases the layer spacing corresponding to a single molecular length, slightly decreasing on cooling, with no anomalies observed at the phase transitions. No layer spacing change suggests that both tilted phases have nearly the same tilt magnitude, the optical uniaxial properties of lower tilted phase point that it has a helical, $SmC_{TB}$-type structure.

In a cell with planar anchoring conditions (HG cell), the light extinction direction remains along the rubbing direction in all smectic phases. This indicates an anticlinic structure of the upper temperature tilted phase - $SmC_A$ phase. The temperature dependence of optical birefringence ($\Delta n$) is consistent with a continuous increase of molecular tilt below the SmA phase (Figure 4), in the tilted smectic phases, the birefringence begins to decrease from the value extrapolated from the critical dependence found in the nematic and smectic A phases. The $SmC_A$-$SmC_{TB}$ phase transition is marked with a slight, step-like increase in $\Delta n$, and a further deviation from the extrapolated value is observed on cooling. Interestingly, in HT cell careful microscopic observations at the $SmC_{TB}$ to SmC phase transition revealed a selective light reflection phenomenon (Figure 4). In ~1K temperature range, colours change from blue to red with increasing temperature, providing the evidence of the helix unwinding. It is worth noticing that the helix unwinding found in this material is similar to the phenomena observed for dimers at SmC$_{TB-SH}$- SmC$_{TB-DH}$ phase transition [14].

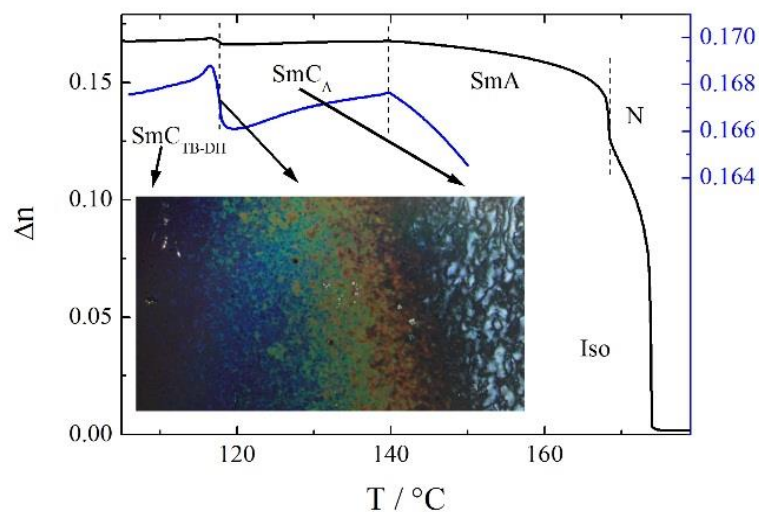

**Figure 4**. Temperature dependence of optical birefringence, Δn (black line), for the homologue **I-8** measured with green light, λ=532 nm. A blue line presents a fragment of zoomed Δn(T) dependence (see right axis). In the inset, the textures observed in HT cell with temperature gradient are presented. In a small temperature range (~1K) near the SmC$_A$- SmC$_{TB-DH}$ phase transition, the colours are observed due to the selective reflection of light from the helical structure formed in the SmC$_{TB-DH}$ phase. The colours change in a sequence blue to red as the temperature increases.

To confirm the identification of SmC$_A$ and SmC$_{TB}$ phases and determine the details of the phase structure, a resonant soft X-ray scattering (RSoXS) experiments were performed for the homologue **I-8** (Figure 5) apart from the standard elastic X-ray diffraction studies. By tuning the incident X-ray beam energy to the edge of the absorption band of carbon (283 eV) it was possible to detect the periodic changes of the molecular orientation. In the SmC$_A$ phase, the signal corresponding to a bilayer periodicity was observed, confirming an anticlinic arrangement of the molecules in the consecutive layers. In the SmC$_{TB}$ phase, the signal symmetrically splits, indicating a helical modulation superimposed on the basic SmC$_A$ phase structure. At 1-2 K below the SmC$_A$-SmC$_{TB}$ phase transition, the helical pitch, estimated from the distance of the split signals in q-space, is ~200 nm (~ 30 smectic layers), unfortunately the crystallisation of the sample precluded more precise determination of the temperature evolution of the pitch length by the resonant x-ray studies. However, the combination of the resonant x-ray and optical studies, in which the selective light reflection from the helix is found in a visible optical range, clearly points to the critical unwinding of the helix upon entering the anticlinic SmC$_A$ phase.

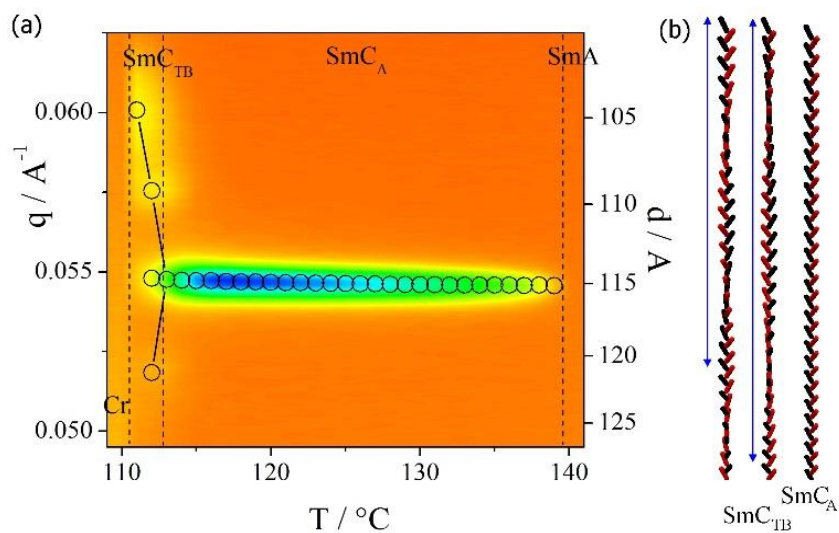

**Figure 5.** (a) Temperature evolution of RSoXS signals for **I-8**, in SmC$_A$ phase the position of the signal corresponds to 2 molecular layers, splitting of the signal in SmC$_{TB-DH}$ phase is caused by additional modulation of the molecular orientation, as the helix is superimposed on the SmC$_A$ bilayer structure. (b) The model of the anticlinic and the helical structures of SmC$_A$ and SmC$_{TB-DH}$ phases, respectively; the structures with two different helical pitch lengths (marked by blue arrows) are presented.

### 2.2. Mesomorphic properties of the long homologues

For homologues with n>9, two smectic phases, SmC$_1$ and SmC$_2$, were found upon cooling below the SmA phase (Figure 2). The XRD studies revealed a layer spacing corresponding to a single molecular length, and a liquid-like in-plane ordering of the molecules in these phases. At the transition SmC$_1$-SmC$_2$, a small step-like increase in the layer thickness was observed. Moreover, the lower-temperature phase exhibited a negative thermal expansion coefficient.

In HG cells, the SmA phase showed a uniform texture, with the extinction direction along the rubbing direction, covered with a regular array of strongly elongated focal conics. A regular array of stripes could be observed in thin HG cells [30-32]. Upon cooling, in the stripe-free regions, several micron-sized domains are formed at the transition to the SmC$_1$ phase, having the extinction direction inclined from the rubbing direction at a few degrees, the domains can be brought into light extinction condition by rotation of the sample either clockwise or anticlockwise (Figure 6a). Such a texture unequivocally indicates a tilted smectic phase. Interestingly, at the SmC$_1$-SmC$_2$ phase transition, the tilt direction reverses within each domain without affecting the domain boundaries; the change of the apparent tilt angle is continuous (Figure 6b). This phenomenon occurs without pronounced changes in the birefringence of the sample (Figure 6a), and it is reversible upon heating and cooling. To confirm that this effect is inherent to the material and not induced by surface interactions, samples of various thickness were tested, the tilt direction inversion was consistently reproduced in all of them.

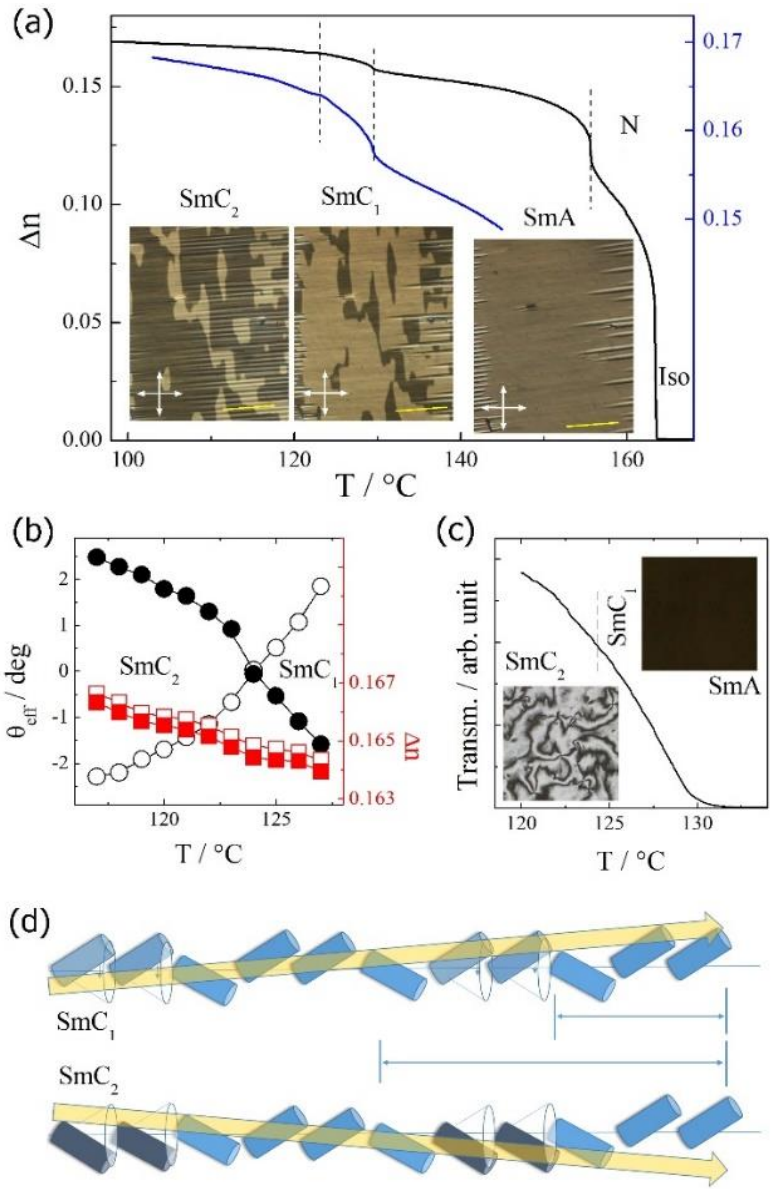

**Figure 6**. (a) The temperature dependence of the optical birefringence, $\Delta n$ (black line) for the homologue **I-10** (blue line presents a fragment of the zoomed $\Delta n(T)$ dependence, see the right axis). In the inset, the textures of the SmA and tilted $SmC_1$ and $SmC_2$ phases, the brightness of tilt domains in the SmC phases is interchanged as the tilt inversion occurs at the $SmC_1$-$SmC_2$ phase transition. Arrows mark polarizer directions, yellow lines show the rubbing direction, slightly inclined from the polarizer direction. (b) the temperature dependence of the apparent tilt angle ($\theta_{eff}$) in the two types of domains (open and solid black circles) as well as the optical birefringence (red squares) measured in μm-size spots in the domains across the $SmC_1$-$SmC_2$ phase transition. (c) The light transmission through the HT cell; the inversion of the apparent tilt at the $SmC_1$-$SmC_2$ phase transition is not connected with any light transmission anomaly, thus, the in-plane birefringence change. In the inset photographs, the same area of the sample is shown in the SmA and $SmC_2$ phases. (d) The model of the tilt arrangement in the consecutive smectic layers in the $SmC_1$ and $SmC_2$ phases; a 3-layer repeating unit (upper blue arrow) of the $SmC_1$ changes to a 6-layer unit (lower blue arrow) in the $SmC_2$ phase by the molecular rotation on the tilt cone in the marked layers, which results in the inversion of the apparent optical tilt direction (yellow arrows).

In the HT cell, both tilted smectic phases exhibit a schlieren texture, confirming their optical biaxial character (Figure 6c). The in-plane birefringence (and, thus, the optical texture) shows no anomaly at the $SmC_1$-$SmC_2$ phase transition, which suggests that the transition is not accompanied with a change in the tilt magnitude. Apparently, none of the tilted phases is simple synclinic or anticlinic type, both have a complex sequence of synclinic (*S*) and anticlinic (*A*) interlayer interfaces. The exact mechanism behind the $SmC_1$-$SmC_2$ phase transition needs to be elucidated, but it surely involves the change in order and number of synclinic and anticlinic interfaces. One of the simplest possibilities involves doubling of the basic multilayer repeating unit on the transition from the $SmC_2$ to $SmC_1$ phase, with the sequence of interlayer interfaces changing from a 3-layer *SAA* unit to a 6-layer *SASSSA* unit (Figure 6d). When observed in a planar cell geometry, such a change results in the inversion of the effective/apparent optical tilt direction, without affecting its magnitude in single layers.

**Mesomorphic properties of the homologue I-9**

The homologue with n=9 displays the most complex phase sequence, featuring five tilted smectic phases below the SmA phase, all characterised with a liquid-like in-plane order of the molecules. Each smectic-smectic phase transition is accompanied with a small, step-like anomaly in the optical birefringence (Figure 7). The optical textures of all the phases observed in HG cells resembled those observed for longer homologues, also the inversion of the apparent tilt angle in tilted domains was found (Figure S4). We decided to label the phases as $SmC_1$', $SmC_1$, $SmC_1$'' and $SmC_2$', $SmC_2$ on cooling from Iso. The inversion phenomenon occurs at the transition between the $SmC_1$'' and $SmC_2$' in analogous way described for the homologues with n>9.

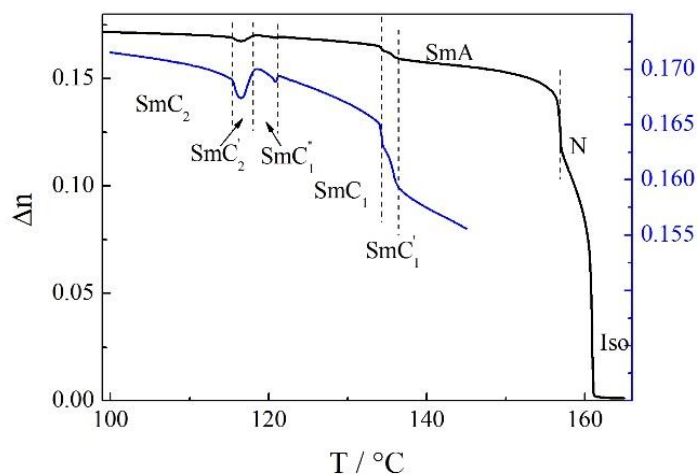

**Figure 7**. The temperature dependence of the optical birefringence for **I-9**. Blue curve presents this dependence in an enlarged view (see the right scale).

The transitions between the smectic phases are not accompanied with a significant change in textures observed in planar geometry. In HT cells, however, the phase transitions are accompanied with changes of optical textures (Figure S5), some of the observed textures are indicating the formation of 2D modulated phases. Below the optically uniaxial SmA phase, a weakly birefringent fan-like texture is formed, and only the lowest $SmC_2$ phase is characterised with a simple schlieren texture, typical for a simple lamellar tilted smectic phase. Similarly to longer homologues, there are no texture changes at the transition between the SmC$_1$ and SmC$_2$ phases, at which the inversion of the apparent tilt occurs.

The X-ray diffraction studies revealed that in all the smectic phases the main periodicity related to the density modulation along the layer normal (indicated with the strongest diffraction signal) closely matches the molecular length. However, in all the phases except the SmA and the lowest temperature SmC$_2$, additional weak signals associated with a bilayer structure are detected (Figure 8a). This finding suggests the breaking of the up-down equality in the molecular arrangement within the layers, possibly due to the interactions between strongly polar nitrophenyl end-groups and the formation of 'weak dimeric' units. In none of those phases the positions of the additional signals were commensurate with the diffraction signals coming from the layer thickness, signifying that the electron density is modulated also along the smectic layers, and 2D-modulated structures are formed, all having the oblique crystallographic unit cells (Figure S6). The determined in-plane modulation period changes in the consecutive phases, in all cases being very long, 200-300 Å. Moreover, the inclination angle of the crystallographic unit cell is different in each modulated phase, and not consistent with the tilt found by optical methods in these phases. Therefore, we conclude that the in-plane density modulations grow independently of the tilt structure, possibly the density waves that define a 2D crystallographic structure develop in the direction perpendicular to the tilt plane (Figure 8b).

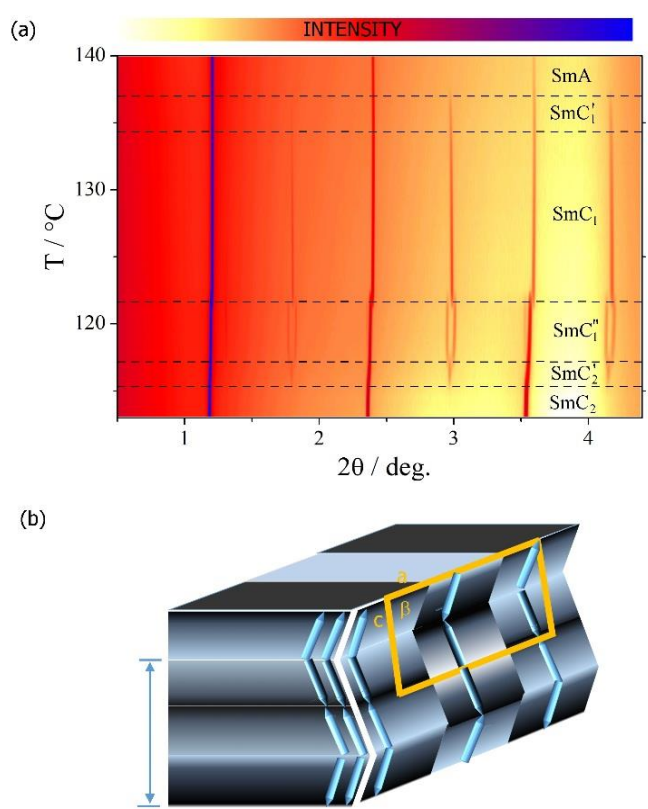

**Figure 8**. (a) The temperature evolution of the small angle X-ray diffraction signals recorded for **I-9** homologue in all the smectic phases. (b) the model of the modulated structure, in which in-plane electron density modulations develop in the direction perpendicular to the tilt plane, a blue arrow shows a 3-layer basic periodicity related to the tilt arrangement, a yellow frame presents a 2D crystallographic unit, parameter c corresponds to a bilayer periodicity. Light/dark grey shadowing indicates the modulation of the electron density due to the breaking of up/down symmetry of the molecules (schematically shown as blue pencils) within the layers.

## Conclusions

The most common tilted smectic phases exhibit either a synclinic or anticlinic arrangement of molecules within consecutive layers. In certain systems, the energy associated with these two packing modes is comparable, resulting in strong frustration that, for some rod-like molecules, might be relieved by building structures with a complex sequence of synclinic and anticlinic interlayers. These structures might display a periodicity extending from 3 up to several layers [33]. Conversely, the competition between the synclinic and anticlinic interactions may also be resolved by building structures with interfaces that are neither synclinic nor anticlinic. Moving from layer to layer, the molecules undergo rotation on the tilt cone at arbitrary angles between 0 and π, leading to the formation of short helices, with clock (SmCα) or distorted clock-like structures – such structures for rod-like molecules are facilitated by molecular chirality [1]. Still another scenario was found for bent dimers, where the interplay between the synclinic and anticlinc interactions leads to the formation of an incommensurate double helical structure [14], in which an additional rotation of the molecular position on a tilt cone is superimposed on a short 4-layer helical unit cell.

Here, we have investigated the molecules with a unique structure, which has not yet been introduced. The presented rigid bent-core molecules have a bulky and polar end group attached through the flexible spacer. Depending on the length of the spacer, the phase behaviour of the material is drastically different - shorter homologues have a strong odd even effect in the transition temperatures, which confirms that the bulky end group is either along the arm or inclined to the arm; for longer homologues the conformational freedom of the spacer makes the transition temperatures less sensible to the spacer length. For all, short and long molecules, the competing anticlinic and synclinic interactions result in the formation of the extended multilayer periodic structures. In the case of short homologues, below the anticlinic $SmC_A$ phase, the bilayer structure adopts a helically modulated configuration ($SmC_{TB}$), optimizing energy by creating double interlocked helices with the layer interfaces that are neither synclinic nor anticlinic but exhibit a greater uniformity across the structure. In contrast, longer homologues resolve competing interactions by establishing multilayer structures with a complex sequence of synclinic and anticlinic interfaces. As the temperature decreases, the ratio between the synclinic and anticlinic interfaces changes, giving rise to an 'apparent tilt' inversion. The specific structure relieving frustration depends on the strength of the interactions between the layers striving to maintain molecules within a single plane. The present work contributes to understanding of the complex structure of fluid phase development, with a particular focus on the emergence of chiral structures from achiral molecules.

## Conflict of interests

The authors declare no conflict of interests.

## Acknowledgements

Authors acknowledge the project FerroFluid, EIG Concert Japan - 9th call "Design of Materials with Atomic Precision". The beamlines 7.3.3 and 11.0.1.2 at the Advanced Light Source at the Lawrence Berkeley National Laboratory are supported by the Director of the Office of Science,